# Simultaneously shaping the intensity and phase of light for optical nanomanipulation

Xionggui Tang,*,†,‡ Fan Nan,§,† Fei Han† and Zijie Yan*,§,†

†Department of Chemical and Biomolecular Engineering, Clarkson University, Potsdam, New York 13699, United States
‡Department of Physics, Key Laboratory of Low Dimensional Quantum Structures and Quantum Control of Ministry of Education, Synergetic Innovation Center for Quantum Effects and Applications, Hunan Normal University, Changsha, 410081, P.R. China
§Department of Applied Physical Sciences, University of North Carolina at Chapel Hill, Chapel Hill, North Carolina, 27599, United States

*E-mail: tangxg@hunnu.edu.cn (X. Tang); zijieyan@unc.edu (Z. Yan)

**Abstract:** Holographic optical tweezers can be applied to manipulate microscopic particles in arbitrary optical patterns, which classical optical tweezers cannot do. This ability relies on accurate computer-generated holography (CGH), yet most CGH techniques can only shape the intensity profiles while the phase distributions are random. Here, we introduce a new method for fast generation of holograms that allows for simultaneously shaping both the intensity and phase distributions of light. The method uses a discrete inverse Fourier transform formula to directly calculate a hologram in one step, in which a random phase factor is introduced into the formula to enable simultaneous control of intensity and phase. Various optical patterns can be created, as demonstrated by the experimentally measured intensity and phase profiles projected from the holograms. The simultaneous shaping of intensity and phase of light provides new opportunities for optical trapping and manipulation, such as optical transportation of metal nanoparticles in ring traps with linear and nonlinear phase distributions.

# 1 Introduction

Since Arthur Ashkin invented the classical optical tweezers (i.e., a tightly focused laser beam),[1] various optical trapping systems have been developed.[2-3] Among them, holographic optical tweezers have attracted much attention due to their several advantages, including dynamic control, high diffraction efficiency, and great flexibility.[2-4] Holographic optical tweezers provide versatile routes for trapping, sorting, transporting, and assembling of micron and nanoparticles.[2,5] In addition, as a general optical and electronic technology, optical holography has been a rapidly growing field in the past two decades. It has found a wide range of applications in laser beam shaping, optical communication, 3D display, biological imaging, microfluidics, atom cooling, and quantum manipulation.[6-16]

A phase-only spatial light modulator (SLM), which can be easily used to generate a desired light field with CGH, has been one of the most attractive devices for building holographic optical tweezers.[17,18] A phase-only SLM consists of a 2D matrix of liquid crystal cells individually controlled by electronic signals. It can retard the optical phase from 0 to $2\pi$, which is determined by the corresponding grey level of the calculated CGH. In the past several years, various methods have been proposed for the design of phase-only holograms, which generally can be categorized as the iterative algorithm, non-iterative algorithm, and integral methods. For most iterative algorithms, such as the iterative Fourier transform algorithm,[19] direct binary search,[20] Gerchburg-Saxton algorithm,[21-24] and adaptive-additive algorithm,[25] the intensity distribution with relatively good uniformity at the output plane can be obtained by iterative searching and optimization. However, the phase distribution is generally random, which is obviously unfavorable for nanoparticle manipulation when using optical forces arising from phase gradients.[26] In addition, the quality of CGH highly depends on the steps and searching algorithm in the optimization process, which largely affects the computation time of hologram generation.

For non-iterative algorithms, the computation time can be reduced, and the speckle noise can be decreased to some extent.[27, 28] However, the quality of the generated hologram still needs to be improved. In order to achieve simultaneous intensity and phase control, Jesacher et. al. proposed a method of using two cascaded phase diffractive elements to separately modulate the amplitude and phase of a laser beam,[29] but it is difficult to precisely control the alignment between two phase elements, which easily results in poor quality of the optical field and low efficiency. Later, Bolduc et. al. established a similar method of encoding the amplitude and phase of the optical field by using two SLMs, which can be applied in the quantum encryption.[30] Recently, Rodrigo et. al. demonstrated a strategy for CGH design of optical traps by using the integral method, which can effectively achieve simultaneous control of intensity and phase distributions.[31-35] However, this method is limited to the realization of smooth optical patterns consisting of special curves, termed as superformula curves. If an optical pattern cannot be expressed by a well-defined formula, the integral method cannot be used directly, and some approximations must be made. Specifically, the method is not suitable for designing even ordinary optical profiles, such as straight lines and point arrays. Therefore, it is highly desirable to generate optical fields with arbitrary control of intensity and phase.

In this work, we introduce a new approach for generating phase-only holograms that can produce high-quality intensity and phase profiles at the output plane. The approach is a direct computation method for CGH. Its simple calculation leads to low computational cost yet high accuracy. Through numerical simulation and experimental investigation, we show that the intensity and phase distribution can be simultaneously obtained, which is very useful for realizing new functions in optical manipulation.

## 2 Method

We assume a reflective pure phase SLM is illuminated by a linearly polarized laser beam with a wavelength of $\lambda$. The phase-modulated beam is at the back focal plane of a lens with a focus length of $f$, generating a holographic optical pattern at its front focal plane. The relationship between optical fields at the input and output planes can be written as[36]

$$H(x_i, y_i) = c \iint U(x_o, y_o) \exp[j \frac{2\pi}{\lambda f}(x_i x_o + y_i y_o)] dx_o dy_o, \qquad (1)$$

where $c$ is a constant, $j$ is complex unit, $H(x_i, y_i)$ is the optical field reflected at the input plane, and $U(x_o, y_o)$ is the desired optical field at the output plane. These optical fields be expressed as,

$$U(x_o, y_o) = a_u(x_o, y_o) \exp[j\varphi_u(x_o, y_o)], \qquad (2)$$
$$H(x_i, y_i) = a_h(x_i, y_i) \exp[j\varphi_h(x_i, y_i)], \qquad (3)$$

Obviously, there exists an inverse Fourier Transform relationship between $H(x_i, y_i)$ and $U(x_o, y_o)$, as shown in Eq. (1). Theoretically, $H(x_i, y_i)$ can be easily obtained by using a fast inverse Fourier Transform (IFFT) if the SLM can modulate both the amplitude and phase. However, commercial SLMs are either amplitude-only or phase-only devices. For example, if the calculated optical field $H(x_i, y_i)$ is projected on a phase-only SLM, all amplitude information $a_h(x_i, y_i)$ will be lost, leading to a large discrepancy between the generated optical field $U'(x_o, y_o)$ and the desired optical field $U(x_o, y_o)$. Usually, the discrepancy will further increase as the bandwidth of amplitude $a_h(x_i, y_i)$ gets wider.

From a different standpoint, the optical field $H(x_i, y_i)$ can be viewed as the superposition of individual plane waves with a weight factor, $U(x_o, y_o)$. Hence, the Eq. (1) can be given using discrete expression,

$$H(m,n) = c' \sum_{\substack{k=-K/2 \\ s=-S/2}}^{\substack{k=K/2 \\ s=S/2}} U(k,s) \exp[j \frac{2\pi}{\lambda f}(km\Delta x_o \Delta x_i + sn\Delta y_o \Delta y_i)], \qquad (4)$$

where $\Delta x_o$ and $\Delta y_o$ are the single pixel size in the $x$ and $y$ direction at the output plane; $\Delta x_i$ and $\Delta y_i$ are the single pixel size in the $x$ and $y$ direction at the input plane; $c' = c\Delta x_o \Delta y_o$; $k \in [-K/2, K/2]$, $s \in [-S/2, S/2]$, $m \in [-M/2, M/2]$, $n \in [-N/2, N/2]$, in which $K$ and $S$ stand for the maximum number of sampling pixels in the $x$ and $y$ axis at the output plane, respectively, and $M$, $N$ denote the maximum number of sampling pixels in the $x$ and $y$ axis at the input plane, respectively; $U(k,s)$ stands for $U(k\Delta x_o, s\Delta y_o)$, and $H(m,n)$ denotes $H(m\Delta x_i, n\Delta y_i)$. Unfortunately, we cannot directly use Eq. (4) to calculate the CGH, as Eq. (4) is mathematically equivalent to Eq. (1). To address this issue, we innovatively modify Eq. (4) by introducing a random phase factor $\exp(j\phi)$, in which $\phi = d_m \cdot rand$, $d_m$ is the modulation depth with a maximum of $2\pi$, and $rand$ is a function that creates a random number in the interval (0,1) in each step of calculation. Hence, the Eq. (4) can be written as the following equation:

$$H(m,n) = c' \sum_{\substack{k=-K/2 \\ s=-S/2}}^{\substack{k=K/2 \\ s=S/2}} U(k,s) \exp[j \frac{2\pi}{\lambda f}(km\Delta x_o \Delta x_i + sn\Delta y_o \Delta y_i) + j\phi]. \qquad (5)$$

In this case, a random phase generated by Matlab library function is directly added into each plane wave used as an eigenfunction in the superposition process; therefore Eq. (5) is substantially different from Eq. (4). The random phase is employed to minimize the crosstalk (i.e., interference) among different plane waves at the input plane, so it effectively improves the beam shaping capacity of the hologram, which leads to the realization of simultaneous control of the intensity and phase profiles. In other words, it is helpful for transferring from the amplitude information to phase information while using the modified discrete inverse Fourier transform, so the calculated phase-only hologram has higher accuracy to generate the designed optical field compared with the one without random phase. Our simulated and experimental results demonstrate that the reconstructed intensity and phase profiles at the output plane are highly consistent with the desired optical pattern.

It is worth noting that the holograms in our proposed method are directly calculated in single steps, which are different from the conventional iterative methods. The Gerchburg-Saxton algorithm may also use random phases to preset the initial phase of a hologram in the first step of the search process, but the purpose is only to converge to the solution faster.[21] Our method has not been demonstrated previously, but a very recent report adopted a similar concept of the random phase factor.[37] The report shows that Fresnel holograms can produce large volume, high-density, dynamic 3D images at different depths by adding random phases into each desired image, which has wide potential for 3D display. However, the iterative Fourier transform algorithm has been used for generating a set of kinoforms, and phase profiles of each image at different output positions cannot be fully controlled.[37] In our method, the random phase factor is directly introduced in each plane wave of Eq. (5) to generate the desired optical traps with intensity and phase profiles, which is highly preferred in the optical manipulation of particles.

## 3 Results

Our proposed method is employed to generate CGH at the input plane and reconstruct the image at the output plane to test its validation. Our target patterns include point spot arrays, peanut-shaped spot arrays, straight-lines, and rings. In the simulation process, each target pattern has 1024×1272 pixels, and its pixel size is 0.139 μm × 0.139 μm at the output plane. Similarly, the phase-only SLM has 1024×1272 pixels with a pixel size of 12.5 μm × 12.5 μm. The focus length of the lens is 3 mm. The related experimental investigations are carried out to further evaluate the proposed method. These experiments were performed in our optical trapping system,[12] which includes a CW tunable Ti:Sapphire laser (Spectra-Physics 3900s) operating at a wavelength of 800 nm and producing a TEM$_{00}$ Gaussian mode. A phase-only SLM (Hamamatsu X13138) is used to modulate the phase of the calculated hologram. A beam profiler (Edmund Optics) is employed for capturing the optical intensity of the reconstructed image, and a Shack-Hartmann wavefront sensor (Thorlabs WFS20-7AR) is used to measure the phase, which has resolution of 1440 x 1080 pixels and arrays of 47 x 35 active lenslets. In addition, a grating phase is added to the calculated hologram to shift the reconstructed image, which can effectively reduce the negative influence from the zero-order optical diffractive spot of SLM.

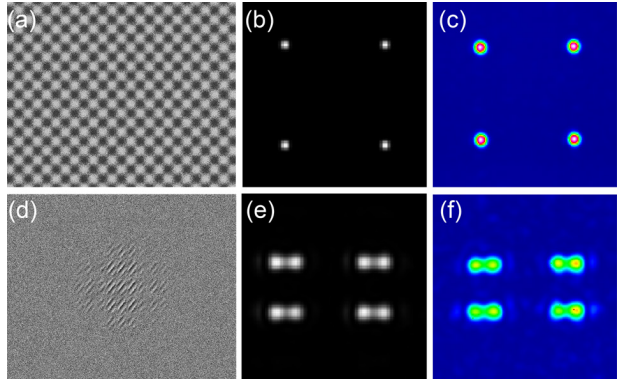

**Figure 1:** Point and peanut-shaped spot arrays. (a, d) Calculated holograms, (b, e) reconstructed intensity profiles, and (c, f) measured intensity profiles.

*Point spot arrays.* The point spot arrays are conventional optical tweezers, which are widely applied in optical trapping, ranging from atoms to microparticles. Figure 1(a, b) shows the calculated hologram and reconstructed image of a point optical trap array. The experimentally measured optical pattern is presented in Figure 1(c), which agrees well with the reconstructed image and demonstrates that the calculated hologram has high accuracy. The intensity distribution among the optical traps is very uniform, which is highly desired for optical trapping.

*Peanut-shaped spot arrays.* The peanut-shaped spot arrays are novel optical traps, which have potential for achieving new functions in optical manipulation.[38] Similarly, the computed hologram, reconstructed intensity, and measured intensity are given in Figure 1(d)~(e), respectively. Again, the intensity profiles of all peanut-shaped spots are nearly identical. In addition, the difference between the simulated results and the experimental results is very small, which further demonstrates that our method can generate high-quality holograms.

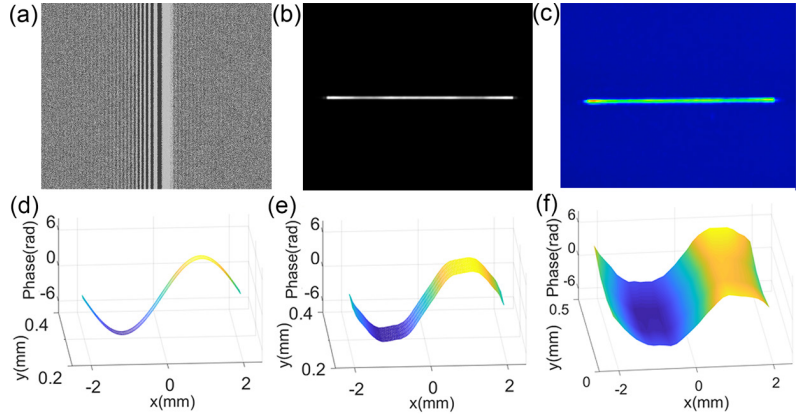

**Figure 2:** Optical line trap. (a) Calculated hologram, (b) reconstructed intensity, (c) measured intensity, (d) designed phase, (e) reconstructed phase, and (f) measured phase.

*Optical line traps.* The optical line traps generated by cylindrical lens holograms have non-uniform intensity and phase distributions,[13, 14] which limits their application in optical manipulation. Flat-top line traps with specified phase profiles can be produced by a shape-phase modulation approach,[39] but the light utilization efficiency is low. Here, we design an optical line pattern, which has uniform optical intensity and phase distribution with a sinusoidal function. The calculated hologram, reconstructed optical intensity, and measured optical intensity are presented in Figure 2(a)~(c), respectively. Evidently, both the reconstructed optical intensity and measured optical intensity from the hologram have uniform distributions, and the former is consistent with the latter. Particularly, the phase distribution with a sinusoidal function in an optical line trap is realized for the first time. The phase distribution is measured at the output plane of a lens with a focus length of 75 cm.[40] In this work, we only measure the phase distribution of the optical line trap due to the limited resolution of the wave-front sensor, which could be inaccurate when measuring the complex wave-fronts of 2D patterns. The designed phase, simulated phase, and measured phase of the optical line trap are given in Figure 2(d)~(f), respectively. The measured phase distribution is in good agreement with the designed and reconstructed phase distributions. The results demonstrate that our method is very helpful for realizing new functions in optical manipulation of nanoparticles by designing different phase distributions.

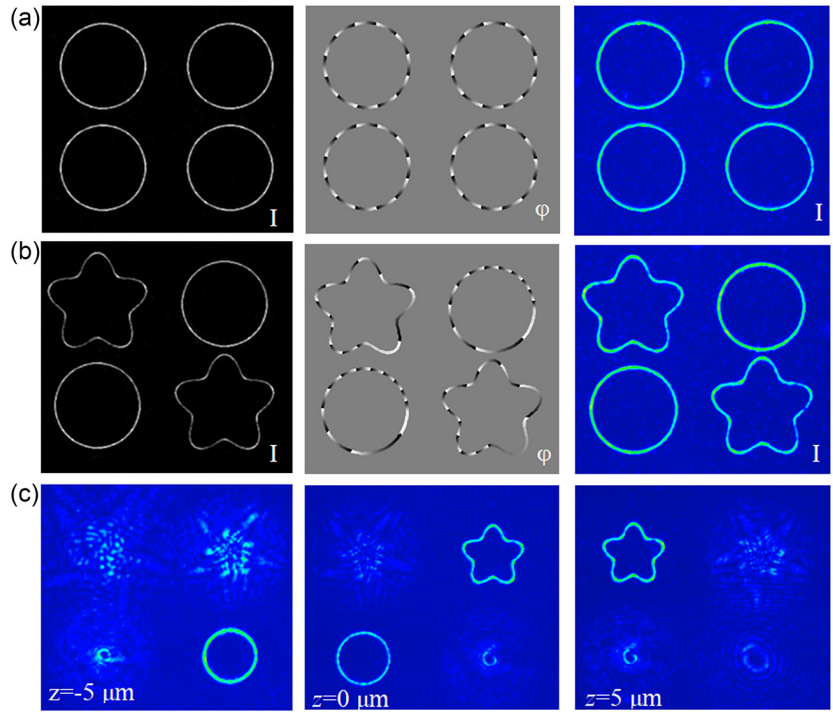

**Figure 3:** Optical trap array of rings and stars. (a,b) Reconstructed intensity (I), reconstructed phase ($\varphi$), and measured intensity of ring trap arrays and mixed ring and star traps. Note that in the middle panel of (b), the phase profiles of the rings and stars are nonlinear. (c) Measured intensities of mixed ring and star trap arrays at different axial positions.

*2D and 3D optical patterns.* The optical trap arrays with closed curves have attracted increasing attention in recent years.[41] For our proposed method, more interestingly, the optical trap arrays can be easily realized by utilizing the convolution theorem from Fourier optics.[36] For simplicity, we present two types of optical traps. The first pattern is four identical rings, whose phase profiles are linear. The reconstructed intensity, reconstructed phase, and measured intensity are presented, as shown in the left, middle, and right subfigures of Figure 3(a), respectively. The second pattern includes two identical rings and two identical stars, whose phase profiles are nonlinear. Similarly, its reconstructed intensity, phase and measured intensity are shown in the left, middle, and right subfigures of Figure 3(b), respectively. Obviously, the difference between reconstructed and measured intensity in each pattern is very small, which indicates that different array patterns can be easily achieved by using our proposed method. More importantly, the desired phase distribution would not be affected by the desired intensity distribution in our method, which is highly preferred in optical manipulation of particles using phase gradient force. Therefore, arbitrary phase distribution can be accurately achieved by using our method. For the iterative algorithms, however, it remains very difficult to precisely generate the desired phase gradient distribution.

Moreover, 3D optical traps with various closed curves at different regions along light propagation direction can also be realized by using our method. The position of each trap can be easily controlled by adding a spherical wave phase with $\pi z(x_o^2 + y_o^2)/(\lambda f^2)$ in the desired hologram, according to on-demand shifting distance $z$ along the optical axis direction. The measured intensity distributions at different positions are shown in Figure 3(c). The ring and star line trap are focused on the focal plane ($z=0$), as presented in the middle subfigure of Figure 3(c), while the other ring and star trap are axially shifted to the plane of $z=-5$ μm and $z=5$ μm, as shown in left and right subfigures of Figure 3(c), respectively. The experimental results are consistent with the simulated results (see Note 1 and Figure S1 in supplementary materials). These multiple traps at different axial positions have important applications for simultaneously implementing multitask optical manipulation.

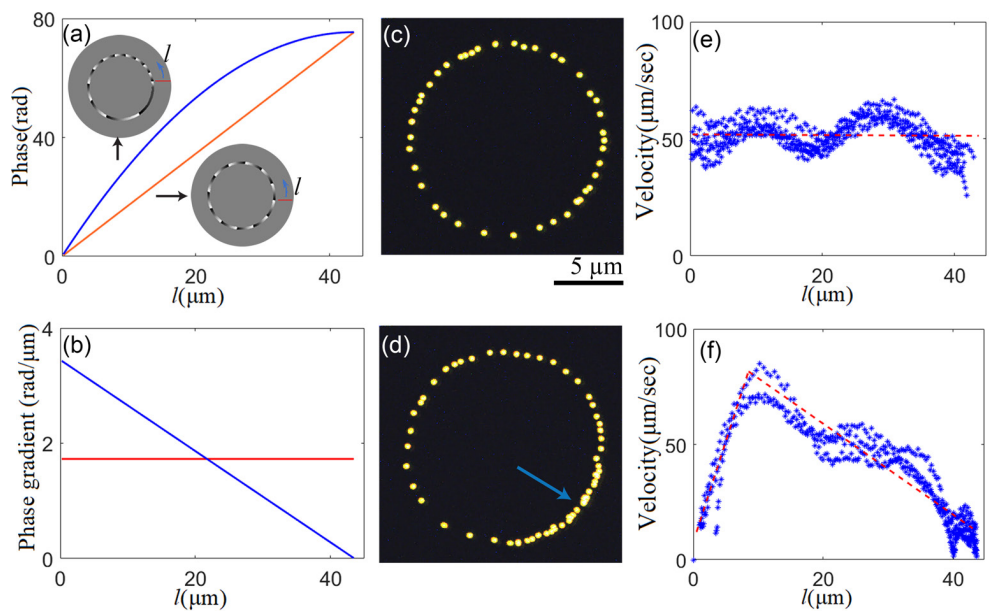

**Figure 4:** Optical manipulation of nanoparticles in ring traps with linear and nonlinear phase profiles. (a) phase variation along circle orbit, (b) phase gradient along circle orbit, (c) superimpositions of dark-field optical images of the trapped nanoparticles moving in a ring trap for one circle with linear phase variation, (d) superimpositions of dark-field optical images of the trapped nanoparticles moving in a ring trap for one circle with nonlinear phase variation, (e) velocity at different positions of a ring trap with linear phase variation, and (f) velocity at different positions of a ring trap with nonlinear phase variation. Note that the insets in (a) are reconstructed phase images, and in (c, d), the time interval from one position of the nanoparticle to its next position is 20 ms.

***Optical manipulation of nanoparticles.*** We performed experiments with Au nanoparticles (200 nm in diameter) to evaluate the performance of optical manipulation in ring traps with different phase distributions. The phase distribution and phase gradient along a circle orbit in the counterclockwise direction are presented in Figure 4(a) and (b), respectively, in which linear and nonlinear phase distributions are denoted by red and blue solid lines, respectively, and the starting points are marked on the reconstructed phase images in the insets of Figure 4(a). A single Au nanoparticle is driven by the phase gradient force along each circle orbit, as shown in the superimposed dark-field optical images in Figure 4(c) and (d), and its speed along the circle orbit is closely related to its phase gradient. The small distance between the two adjacent positions indicates lower velocity (e.g., in the portion marked by the arrow in Figure 4(d), where the phase gradient is small). The measured velocities around the circle orbits are shown in Figure 4(e) and (f), respectively. For the ring optical trap with linear phase variation, its velocity fluctuates moderately around 51 μm/s, as marked by a dashed red line in Figure 4(e). The variation of velocity is due to the anisotropy of optical force in the ring trap created by polarization-dependent focusing of a linearly polarized beam.[42] If a radially polarized or azimuthally polarized beam is employed in the ring trap, the anisotropy effect could be largely eliminated. For the ring optical trap with nonlinear phase variation, its velocity rapidly increases in a short distance and then decreases slowly in a long distance. In the beginning, the velocity is very low, but its phase gradient force is large, so its driving force is much larger than its drag force, which results in rapid increase of velocity. After a short distance, its phase gradient force quickly decreases, so it is smaller than its drag force, leading to a decrease in velocity. Its tendency of velocity variation is marked by the dashed red line in Figure 4(f). Consequently, it reveals that our designed ring trap has excellent capability of trapping and steering metal nanoparticles in the circle orbit.

*Further discussion of the method.* Although the generated holograms by our method generally have slight phase noise caused by the introduction of a random phase factor, the reconstructed intensity and phase profiles are not substantially affected by the phase noise (see Note 2 and Figure S2 in supplementary materials). However, a relatively large difference between the obtained pattern and target pattern would appear if its amplitude information had a large bandwidth (i.e., a complex shape). To address this issue, pre-compensation can be employed to amend the intensity distribution of the target pattern in advance, which can effectively improve the quality of the obtained pattern (see Note 3 and Figure S3 in Supporting Information). Moreover, it is worth noting that the modulation depth, $d_m$, of the random phase factor should be reasonably chosen. If the modulation depth is too small, the crosstalk between different plane waves at the input plane will not be effectively eliminated. However, if it is too large, the noise in the hologram will have a considerable negative effect on the optical pattern at the output plane. An example of line patterns under different modulation depths is given in Figure S4 in Supporting Information. It appears that when the random phase factor is π, its intensity and phase profiles are the most acceptable.

The computation cost is very low in our method. As a comparison, we also calculate the hologram of a typical ring trap by using the integral method[43] (Figure S2(a) in supplementary materials). The calculated result is almost the same with that using our method, but its computation time is about seven times longer than that of our method.

## 4 Conclusions

In this work, we have proposed a novel approach for realizing phase-only holograms. The approach is a direct method that uses a discrete inverse Fourier transform formula, in which a random phase factor is introduced. The numerical simulation and experimental work show that high-quality holograms can be achieved using our method. In particular, the desired phase distribution is not affected by the desired intensity distribution, which is useful to accurately obtain a preferred phase distribution. In addition, our optical trapping experiments indicate that the holograms generated by our method are suitable for optical manipulation of nanoparticles. Our approach has several advantages, including simple calculation, low computation cost, high accuracy, and high flexibility for generating different types of patterns. Our method also has several potential applications in optical trapping, 3D microfabrication, and laser beam shaping.

**Supporting Information**
Effects of phase noise on reconstructed intensity and phase, and reconstructed intensity and phase profiles at different axial positions.

**Acknowledgements**
This work was supported by the W. M. Keck Foundation. X.T. thanks the Foundation of China Scholarship Council for a visiting scholar fellowship.

**Conflict of Interest**
The authors declare no competing financial interest.

Supporting Information

**Simultaneously shaping the intensity and phase of light for optical nanomanipulation**

Xionggui Tang,[*] Fan Nan, Fei Han and Zijie Yan[*]

**Note 1. Reconstructed intensity and phase profiles in multiple planes of a 3D pattern**
Mixed traps can be achieved at different positions along the optical axis. In Figure S1, the reconstructed intensities and phases of rings and stars at different axial positions are presented. In Figure S1(b, e), the intensity and phase profiles of the bottom-left ring and up-right star can be well generated in the focal plane (z=0), while the intensity and phase profiles of the down-right ring and up-left star can be realized at the plane of z=−5 µm and z=5 µm as given in Figure S1(a, d) and (c, f), respectively. Consequently, different multiple optical traps can be obtained at different axial positions, which are helpful for multitask optical trapping.

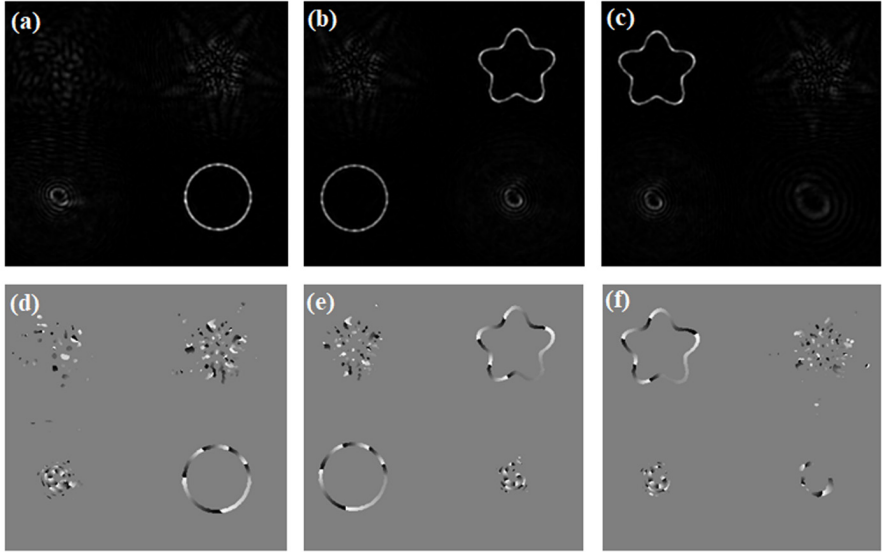

**Figure S1:** A 3D optical trap of mixed rings and stars at different positions along the axial direction. (a, b, c) reconstructed intensity profiles, and (d, e, f) reconstructed phase profiles. The position of (b, e) is at the focal plane (z=0), and the positions of (a, d) and (c, f) are at the plane of z=−5 µm and 5 µm, respectively. Note that all phase profiles are nonlinear.

**Note 2. Effects of phase noise on reconstructed intensity and phase profiles**
In our method, the introduction of random phase factor is helpful for weakening crosstalk among plane waves in order to generate a desired optical field at the output plane. In this case, the phase factor will induce phase noise on the generated hologram. Here we discuss the effect of phase noise on the reconstructed intensity and phase profiles. We assume that the phase distribution of an ideal hologram is expressed as $P_1(x_i, y_i)$, and the phase distribution of generated hologram by our method is written by $P_2(x_i, y_i)$. The phase noise is regarded as $P_3(x_i, y_i)$, which originates from the random phase factor employed in Eq. (5). As a result, the relationship among them is given below,

$$P_2(x_i, y_i) = P_1(x_i, y_i) + P_3(x_i, y_i). \tag{S1}$$

Similarly, the idea amplitude and reconstructed amplitude at the output plane are written by $A_1(x_o, y_o)$ and $A_2(x_o, y_o)$, and the ideal phase and reconstructed phase at the output plane are expressed by $P_{11}(x_o, y_o)$ and $P_{22}(x_o, y_o)$. By using Fourier transform, we can obtain the following formulas

$$A_1(x_o, y_o)\exp(jP_{11}(x_o, y_o)) = FT[\exp(jP_1(x_i, y_i))], \qquad (S2)$$

$$A_2(x_o, y_o)\exp(jP_{22}(x_o, y_o)) = FT[\exp(jP_2(x_i, y_i))], \qquad (S3)$$

where *FT* denotes Fourier transform. From the Eq. (S1), (S2) and (S3), we can get

$$A_2(x_o, y_o)\exp(jP_{22}(x_o, y_o)) = FT[\exp(jP_1(x_i, y_i) + jP_3(x_i, y_i))], \qquad (S4)$$

i.e.,

$$\begin{aligned} &A_2(x_o, y_o)\exp(jP_{22}(x_o, y_o)) \\ &= FT[\exp(jP_1(x_i, y_i))] * FT[\exp(jP_3(x_i, y_i))], \end{aligned} \qquad (S5)$$

where $*$ stands for convolution. As we know, $P_3(x_i, y_i)$ is a weak phase noise caused by the introduction of random phase factor, so we have

$$\exp(jP_3(x_i, y_i)) \approx 1 + jP_3(x_i, y_i). \qquad (S6)$$

The frequency width of $P_3(x_i, y_i)$ is very narrow, which can be considered to be proximately equal to the Dirac delta function $\delta(x_o, y_o)$. Consequently, we have

$$A_2(x_o, y_o)\exp(jP_{22}(x_o, y_o)) \approx FT[\exp(jP_1(x_i, y_i))], \qquad (S7)$$

Comparing with eq. S2, we get

$$A_2(x_o, y_o) \approx A_1(x_o, y_o), \qquad (S8)$$

$$\exp(jP_{22}(x_o, y_o)) \approx \exp(jP_{11}(x_o, y_o)). \qquad (S9)$$

It demonstrates that the phase noise induced by random phase factor has negligible negative affect on the reconstructed intensity and phase.

We provide an example to further verify that our analysis above is reasonable. Here, the holograms, amplitudes and phases of ring traps with and without phase noise are shown in Figure S2. These images show that the intensity and phase reconstructed from the hologram generated by our method with phase noise are nearly identical to those from the ideal hologram without phase noise. The Fourier transform of phase noise is almost equal to the Dirac delta function, which reveals that the phase noise has no substantial influence on the reconstructed intensity and phase. We thus conclude that reasonably accurate intensity and phase can be reconstructed from the hologram generated by our method.

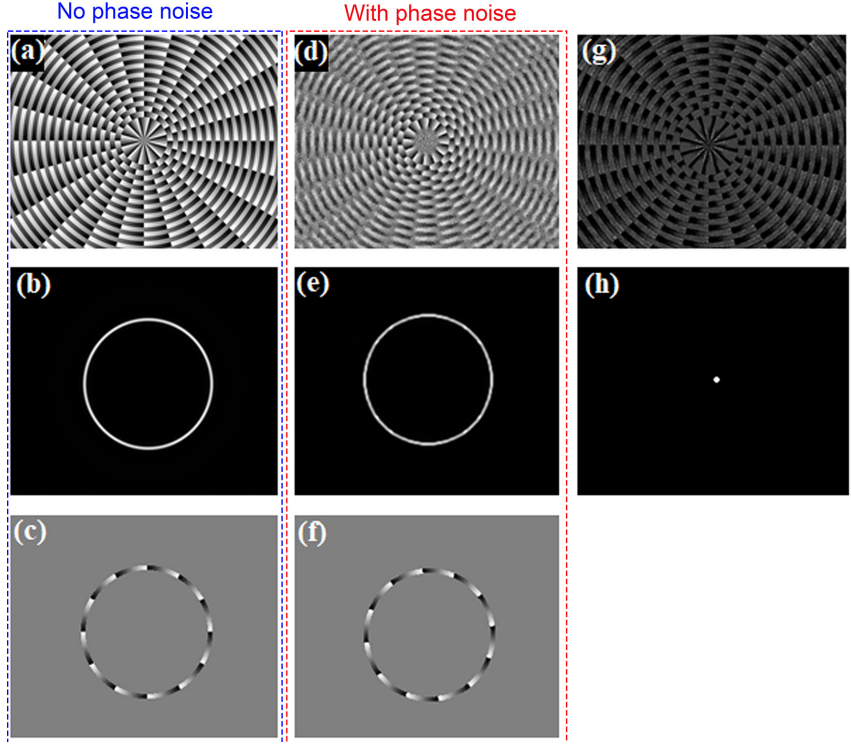

**Figure S2:** Ring traps with linear phase profiles. (a) Hologram for generating an ideal ring trap calculated using the method introduced by Roichman and Grier,[s1] (b) intensity profile reconstructed from the ideal hologram, and (c) phase profile reconstructed from the ideal hologram. (d) Hologram with phase noise calculated by our method, (e) intensity profile reconstructed from our hologram, (f) phase profile reconstructed from our hologram, (g) the phase noise, and (h) Fourier transform of the phase noise.

**Note 3. Optimization of optical patterns**

If the reconstructed optical pattern is relatively poor due to a complex shape, we can pre-compensate its intensity distribution at the weak positions of the target pattern to achieve a more uniform pattern. Here, a rectangle pattern is shown as an example in Figure S3. The intensity of target pattern is uniform, but its reconstructed intensity strongly fluctuates along its orbit as shown in Figure S3(a). According to its profile feature, we simply modify $U(k,s)$ by employing $U(k,s)/(k^2+s^2)$ in Eq. (5), i.e., pre-compensating its intensity profile. The improved intensity distribution is presented in Figure S3(b). The results show that its intensity quality is obviously improved by the optimization process. Specially, its phase profile is not affected by our optimization, which is very helpful for optical manipulation. In addition, the modulation depth, $d_m$, of the random phase factor should be reasonably chosen, and an example is given in Figure S4.

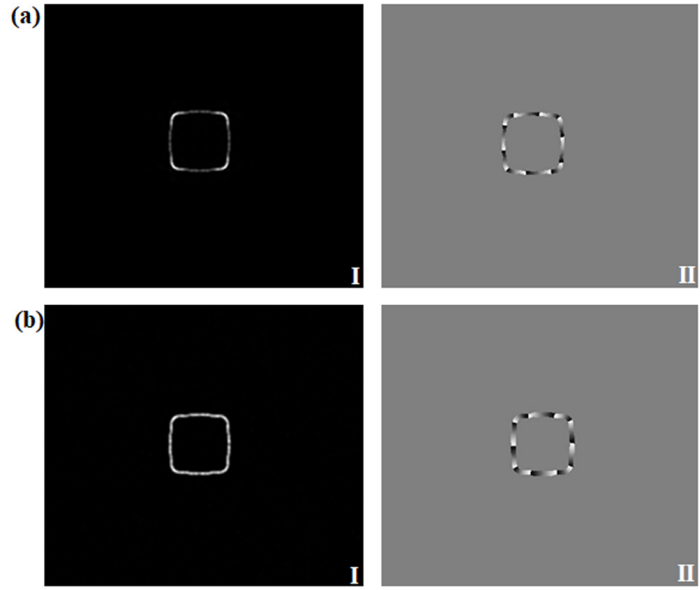

**Figure S3:** Rectangle optical patterns (a) without optimization and (b) with optimization. The patterns in column I denote intensity profiles, and patterns in column II denote phase profiles.

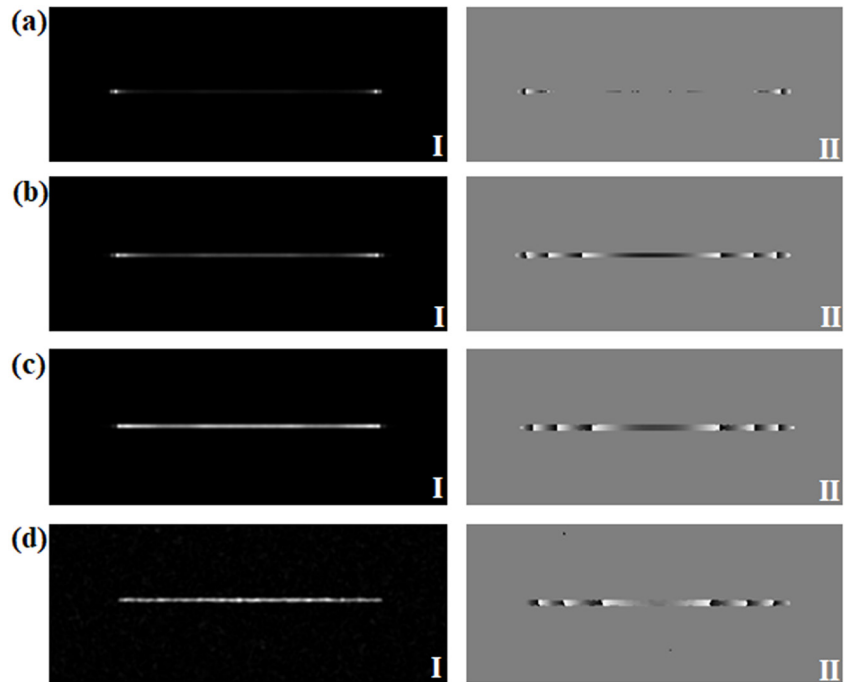

**Figure S4:** Line traps under different modulation depths of random phase. (a) 0, (b) 0.5 π, (c) π, (d) 1.9 π. The patterns in column I denote intensity profiles, and patterns in column II denote phase profiles.